\documentclass[showpacs,preprintnumbers,amsmath,amssymb,aps, prd]{revtex4}
\usepackage{graphicx}
\usepackage{grffile}
\usepackage{mathrsfs, mathtools}
\usepackage{caption}
\usepackage{float}
\usepackage{xcolor}
\usepackage{changepage}
\usepackage{dcolumn}
\usepackage{bm}
\usepackage{graphicx,subfigure,epsfig}
\usepackage{multirow}

\def\a{\alpha}

\def\f{\frac}

\def\e{\mathcal{E}}
\def\l{\mathcal{L}}
\def\c{\cite}

\def\grr{g_{rr}}
\def\gtt{g_{tt}}
\def\gth{g_{\phi\phi}}
\def\veff{\mc{V}_{\text{eff}}}
\def\fr{f(r)}
\def\gr{g(r)}
\def\hr{h(r)}
\def\s{Schwarzschild }
\newcommand\be{\begin{equation}}
\newcommand\ee{\end{equation}}
\newcommand\ba{\begin{eqnarray}}
\newcommand\ea{\end{eqnarray}}
\newcommand\nn{\nonumber}
\newcommand\lt{\left}
\newcommand\rt{\right}
\newcommand\pt{\partial}
\newcommand\tx{\text}
\newcommand\mc{\mathcal}
\begin{document}
\title{Constrain from shadows of $M87^*$ and $Sgr A^*$
and quasiperiodic oscillations of galactic microquasars on a black
hole arising from metric-affine bumblebee model}
\author{Sohan Kumar Jha}
\email{sohan00slg@gmail.com}
\affiliation{Chandernagore College, Chandernagore, Hooghly, West
Bengal, India}
\author{Anisur Rahaman}
\email{manisurn@gmail.com(corresponding author)}
\affiliation{Department of Physics,
Durgapur Government College, Durgapur, Burdwan 713214, West
Bengal, India.}
\date{\today}
\begin{abstract}
\begin{center}
Abstract
\end{center}
We examine a static  spherically symmetric black hole metric that
originates from the vacuum solution of the traceless metric-affine
bumblebee model in which spontaneous Lorentz symmetry-breaking
occurs when the bumblebee fields acquire a non-vanishing vacuum
expectation value. A free Lorentz-violating parameter enters into
the basic formulation of the metric-affine bumblebee model. In
this study, we use observations from the Event Horizon Telescope
(EHT) collaboration on $M87^*$ and $SgrA^*$ to analyse the shadow
of the black hole and an attempt has been made to constrain that
free Lorentz-violating parameter. We also investigate particle
motion over time-like geodesics and compute the corresponding
epicyclic frequencies. We further constrain the Lorentz-violating
parameter by using the reported high-frequency quasi-periodic
oscillations (QPOs) of microquasars, offering new insights into
its possible impact on astrophysical phenomena.

\end{abstract}
\maketitle \textbf{Keywords}: Shadow of black hole, $M87^*$ and
$Sgr A^*$, Metric-affine bumblebee,
 Quasiperiodic Oscillation, Microquasars
\section{Introduction}
In theoretical physics, the quest for a unified understanding of
fundamental forces often lead to the introduction of free
parameters that allow for flexibility in model formulation. Among
these, the Lorentz symmetry-violating parameters are particularly
significant. Lorentz symmetry, a cornerstone of both special
relativity and the Standard Model of particle physics dictates
that the laws of physics remain invariant under transformations
such as rotations and boosts. However, theories that explore
physics beyond the Standard Model, such as quantum gravity models
and extensions of general relativity, often introduce parameters
that could violate this symmetry \cite{COST1, COST2, COST3,
COST4}.

Contracting these free parameters is crucial for several reasons.
First, it helps in constraining the parameter space by aligning
theoretical predictions with experimental observations. For
instance, the Lorentz violation parameter can impact a range of
physical phenomena, from the behavior of particles in high-energy
collisions to the propagation of light in astrophysical contexts
\cite{SAMUEL}. Second, understanding the bounds of such parameters
aids in testing the robustness of theoretical frameworks against
experimental data. By setting limits on the degree of Lorentz
violation, we ensure that new theories remain consistent with
well-established principles and empirical evidence.

In practical terms, contracting Lorentz symmetry-violating
parameters are essential for refining theoretical models and
guiding experimental searches. This process helps to validate or
refute proposed theories while offering insights into the
fundamental structure of spacetime, potentially signaling the need
for new physics beyond the Standard Model \cite{COST1, COST2,
COST3, COST4}.

In the context of black hole physics, investigating Lorentz
violation becomes particularly compelling. Black holes provide a
unique testing ground for probing fundamental physics in extreme
gravitational environments. The introduction of Lorentz-violating
parameters can lead to observable deviations in phenomena such as
black hole shadows \cite{eht, eht1, eht2, eht3}, Hawking radiation
\cite{HAWKING, BAKE}, and quasinormal modes (QNMs)\cite{REGE,
PRESS, VISH, KOKKO, HPN} These deviations are critical as they
help to constrain the magnitude of Lorentz violation by comparing
theoretical predictions with astrophysical observations, such as
the  imaging of black hole shadows by EHT \cite{eht, eht1, eht2,
eht3} or the analysis of gravitational waves from the black hole
mergers \cite{LIGO}.

Black holes, arising as solutions to Einstein's equations, have
long been at the center of understanding the nature of spacetime
and the limits of physical laws. With the discovery of phenomena
such as gravitational lensing \cite{LENSING1, LENSING2, LENSING3}
black hole shadows \cite{eht, eht1, eht2, eht3}, Hawking radiation
\cite{HAWKING, BAKE}, and quasinormal modes (QNMs)
\cite{REGE,PRESS,VISH, KOKKO, HPN}, the study of black holes has
advanced significantly. Recent discoveries, including the
first-ever image of a black hole shadow captured by the Event
Horizon Telescope collaboration \cite{eht, eht1, eht2, eht3} and
the detection of gravitational waves by \cite{LIGO}, have provided
new avenues for investigating black hole properties.

Symmetry plays a fundamental role in theoretical physics, with
Lorentz symmetry lying at the foundation of both the Standard
Model and general relativity (GR). However, it may break at higher
energy scales, as suggested by cosmic ray evidence \cite{COST1,
COST2, COST3} and unified gauge theories \cite{SAMUEL}. Signals of
Lorentz violation at lower energies offers experimental
opportunities \cite{SAMUEL}. Theories like loop quantum gravity
and the Standard Model Extension (SME) \cite{COST1, COST2, COST3}.
accommodate Lorentz symmetry breaking. In particular,
Einstein-bumblebee gravity \cite{COST4} introduces spontaneous
Lorentz symmetry breaking via a bumblebee vector field. Black hole
solutions in bumblebee gravity \cite{CASANA} have led to insights
into phenomena such as Hawking radiation \cite{KANZI} and
traversable wormholes \cite{SAKIL}.  Within the recent few years,
several studies have been made to study the effect of Lorentz
violation of different physical systems and contracting the Lorentz
violation parameter from  different compatible observations
\cite{MALUF, GUIO, ESCO, COSMOLOGY, FANG, MARIZ, ADS, KHODADI}.
 Bumblebee
gravity introduces modifications to spacetime that manifest in
various astrophysical observables, such as gravitational lensing,
black hole shadows, Hawking radiation, QNMs, and quasi-periodic
oscillations (QPOs) \cite{QPO1,QPO2,QPO3, QPO4, QPO5, QPO6, QPO7,
QPO8, QPO9, QPO10}.  The metric-affine Bumblebee gravity
framework, which treats the metric and affine connections
independently, has provided new insights \cite{GHIL, GHIL1}. It is
also, an important model to study the effect of Lorentz violation.
Recent studies \cite{ ADEL, ADEL1, ADEL2, AFFINE, AFFINE1} have
addressed Lorentz symmetry-breaking (LSB) research significantly,
offering new solutions and exploring the effect of the LSB on
light deflection and perihelion advance of Mercury. In
\cite{AFFINEOUR} effects of Lorentz violation on quadrinomial
modes have been studded and an attempt has been made to constrain
Lorentz violation parameter from the observation of $M87^*$
sipermassive black hole.

Hawking radiation, a quantum mechanical effect expected from black
holes, depends crucially on the structure of spacetime near the
event horizon. The Lorentz-violating parameter modifies the
horizon geometry, altering the radiation spectrum and Hawking
temperature. Bumblebee gravity, in particular, leads to deviations
from the thermal radiation profile predicted by GR. These
deviations may be detectable in future observations of black hole
thermodynamics, offering a novel way to test and constrain
Lorentz-violating effects \cite{KANZI}.

QNMs \cite{REGE, PRESS,VISH, KOKKO, HPN, RKONO,CARDOSO, LBAR,
KONO} describe the response to the characteristic ringdown frequencies of
of black holes to perturbations. Since these oscillations are
closely tied to the curvature and geometry of the surrounding
spacetime, the bumblebee parameter introduces shifts in both the
real
 and imaginary parts of the QNM spectrum
\cite{KOKKO, HPN, RKONO, CARDOSO, KONO}.
 These modifications are highly
sensitive to the background metric and offer a direct means of
testing deviations from classical GR. Gravitational wave
observations from LIGO scientific, VRIGO collaboration provide a
promising ground for exploring Lorentz violation in strong-field
regimes.

One of the most direct effects of Lorentz violation in Bumblebee
gravity can be observed through the deflection of light around
compact objects, influencing gravitational lensing \cite{LENSING1,
LENSING2, LENSING3} and black hole shadows \cite{eht, eht1, eht2,
eht3}. Modifications to the spacetime geometry caused by the
Lorentz-violating parameters distort photon trajectories,
potentially leading to observable deviations in the size and shape
of the black hole's shadow. These distortions are especially
relevant in light of precise observations from the Event Horizon
Telescope, which captured the shadows of supermassive black holes
like $M87^*$ and $SgrA^*$ \cite{eht, eht1, eht2, eht3}. Comparing
these observations with Bumblebee gravity predictions allow for
potential constraints on the Lorentz violation parameter.

Finally, a promising observational window for constraining the
bumblebee parameter lies in the study of \cite{QPO1,QPO2,QPO3,
QPO4, QPO5, QPO6, QPO7, QPO8, QPO9, QPO10}, which are oscillatory
features in the X-ray power spectra of black holes and neutron
star systems. These oscillations are tied to the motion of matter
in the accretion disk and are influenced by the underlying
spacetime geometry. In Bumblebee gravity, QPO frequencies are
modified due to the altered metric structure introduced by the
Lorentz-violating vector field. By comparing observed QPO
frequencies with theoretical predictions in the context of
Bumblebee gravity, stringent constraints on the Lorentz-violating
parameter can be established \cite{QPO1,QPO2,QPO3, QPO4, QPO5,
QPO6, QPO7, QPO8, QPO9, QPO10}.

The effects of Lorentz violation in Bumblebee gravity extend
beyond classical tests of general relativity, potentially
reflecting deeper quantum gravitational phenomena. This interplay
between observable astrophysical phenomena and the underlying
quantum structure of spacetime provides a promising avenue for
constraining the Bumblebee parameter. Observational data from
black hole shadows, QNMs, and QPOs enhance our ability to probe
Lorentz symmetry-breaking effects and test quantum gravity
theories \cite{ADEL, ADEL1,ADEL2, AFFINE, AFFINE1, AFFINEOUR}.

Thus, the Lorentz violation parameter in metric-affine
 bumblebee gravity is expected to induce modifications across a range of black
 hole and strong-field phenomena, including gravitational
 lensing \cite{LENSING1, LENSING2,LENSING3}, shadows \cite{eht, eht1, eht2, eht3},
 thermodynamics \cite{HAWKING, BAKE, KANZI}, QNMs  \cite{REGE, PRESS, VISH, KOKKO, HPN,
 RKONO, LBAR, CARDOSO, KONO}
  and QPOs \cite{QPO1, QPO2, QPO3, QPO4, QPO5, QPO6, QPO7, QPO8, QPO9, QPO10}.
  Among these, QPOs stand out as a
particularly effective tool for constraining the Bumblebee
parameter associated with metric-affine Bumblebee gravity due to
their sensitivity to spacetime structure. This interdisciplinary
approach, combining astrophysical observations with theoretical
predictions, provides a robust framework for testing
Lorentz-violating theories and their implications for quantum
gravity.

The rest of the paper is organized as follows. In sec. II we have
given a brief description of the metric-affine traceless bumblebee
mode. Sec. III is devoted to the estimation of Lorentz violating
parameter, which has been carried out using observable from shadow
$M87^*$ supermassive black hole. In Sec. IV contains two
subsections. In Subsection IV-A  we describe the motion of the
particle in a time-like geodesics and compute the epicyclic
frequencies analytically. Subsection IV-B an attempt has been made
to constrain the Lorentz violating parameter by utilizing the
observational results of QPOs for microquasars. Sec. V contains
concluding remarks on the article.

\section{Description of the metric-affine  traceless  bumblebee mode}
The metric-affine (Palatini) formalism is a prevalent framework in
the study of modified gravity theories. Unlike the traditional
metric approach, this formalism treats the metric and the affine
connection as independent dynamical variables, allowing for
greater generality in exploring the structure of spacetime. In
their work \cite{AFFINE}, the authors investigate the traceless
metric-affine Bumblebee model, which incorporates spontaneous
Lorentz symmetry breaking. They derive a static, spherically
symmetric vacuum solution under this framework. The action for
this model is presented in \cite{AFFINE}.
\begin{eqnarray}
\mathcal{S}_{B}&=&\int
d^{4}x\,\sqrt{-g}\left[\frac{1}{2\kappa^2}\left(\mathcal{R}(\Gamma)
+\xi\left(\mathcal{B}^{\mu}\mathcal{B}^{\nu}-\frac{1}{4}\mathcal{B}^{2}g^{\mu\nu}\right)
\mathcal{R}_{\mu\nu}(\Gamma)\right)
-\frac{1}{4}\mathcal{B}^{\mu\nu}\mathcal{B}_{\mu\nu}-\right.\nonumber\\
&-&\left.V(\mathcal{B}^{\mu}\mathcal{B}_{\mu}\pm b^{2})\right]+ \int
d^{4}x\sqrt{-g}\mathcal{L}_{mat}(g_{\mu\nu},\psi), \label{bumb}
\end{eqnarray}
where $\mathcal{B}_{\mu}$ is the bumblebee field, $g^{\mu\nu}$ is
traceless metric, and $V(\mathcal{B}^{\mu}\mathcal{B}_{\mu}\pm
b^2)$ is the potential that spontaneously breaks the Lorentz
symmetry when $b^2=b_{\mu}b^{\mu}$ is a real positive constant.
The potential is assumed to have a minimum at $V'(b_{mu}b^{mu})=0$
and $\mathcal{B}^{\mu}\mathcal{B}_{\mu}\pm b^2=0$ ensuring the
breaking of $U(1)$ symmetry. In this scenario, the bumblebee field
acquires a nonzero vacuum expectation value
$<\mathcal{B}_{\mu}>=b_{\mu}$  Additionally, it is assumed that
the potential reaches zero at its minimum. The  same algebraic
manipulation and the assumptions mentioned above served the
foundation  to derive a static, spherically symmetric metric
\cite{AFFINE}
\begin{equation}
ds^2=-\frac{\left(1-\frac{2M}{r}\right)}{\sqrt{\left(1+\frac{3\alpha}{4}\right)
\left(1-\frac{\alpha}{4}\right)}}dt^2+\frac{dr^2}{\left(1-\frac{2M}{r}\right)}
\sqrt{\frac{\left(1+\frac{3\alpha}{4}\right)}{\left(1-\frac{\alpha}{4}\right)^3}}
+r^{2}\left(d\theta^2 +\sin^{2}{\theta}d\phi^2\right),
\label{metric}
\end{equation}
where $\alpha$ is the Lorentz-violating parameter. In the limit
$\alpha\rightarrow 0$, the Lorentz symmetry breaking (LSB) metric
[\ref{metric}] reduces to the Schwarzschild metric. Additionally,
there is a noticeable difference between the line elements
associated with the bumblebee and metric-affine bumblebee gravity.
In the former one the coefficient of the spacial part only modifies
however in the metric-affine bumblebee both spatial and temporal
part are modified with different factors containing the Lorentz
violation parameter. The Kretschmann scalar invariant
corresponding to this metric reads
\begin{eqnarray}
\mathcal{K} &=& \mathcal{R}_{\lambda\eta\mu\nu}\mathcal{R}^{\lambda\eta\mu\nu}\nonumber \\
 &=&\frac{1}
{r^{6}(4+3\alpha)^{3/2}}
[48 \alpha Mr\sqrt {4+3\alpha}+32M\alpha r\sqrt {4-\alpha}\nonumber \\
\nonumber
 &-& 12 M{\alpha}^{2}r\sqrt {4-\alpha}+32 {r}^{2}\sqrt {4+
3\alpha}+192{M}^{2}\sqrt {4+ 3\alpha}\nonumber \\
 &-& 32{r}^{2}\sqrt {4-\alpha}-16{r}^{2} \alpha\sqrt {4-\alpha}
-12{\alpha}^{2}Mr\sqrt {4+3\alpha} \nonumber \\
&+& 6 {r}^{2}{\alpha}^{2}\sqrt {4-\alpha}+64Mr\sqrt {4-\alpha}-144
\alpha{M}^{2} \sqrt {4+3\alpha}\nonumber \\
&-& 3{M}^{2}{\alpha}^{3}\sqrt {4+3 \alpha}+36
{M}^{2}{\alpha}^{2}\sqrt {4+ 3\alpha}+3 {\alpha}^{2}
{r}^{2}\sqrt {4+ 3\alpha}\nonumber \\
&+&{\alpha}^{3}Mr\sqrt {4+3\alpha}-64Mr\sqrt
{4+3\alpha}-\frac{1}{4}{\alpha}^{3}{r}^{2}\sqrt {4+3\alpha}].
\label{KREST}
\end{eqnarray}
The expression for the Kretschmann scalar invariant (\ref{KREST})
demonstrates that the effects of Lorentz symmetry breaking, as
represented by the parameter $\alpha$, cannot be entirely absorbed
for by mere re-scaling of the coordinates.  When $\alpha \to 0$
we obtain the anticipated standard result corresponds to the
Schwarzschild metric $\mathcal{K}_{S}=\frac{48 M^2}{r^6}$.
\section{Parameter estimation using  observable from shadow}
\begin{figure}[H]
\centering
\includegraphics[width=0.4\columnwidth]{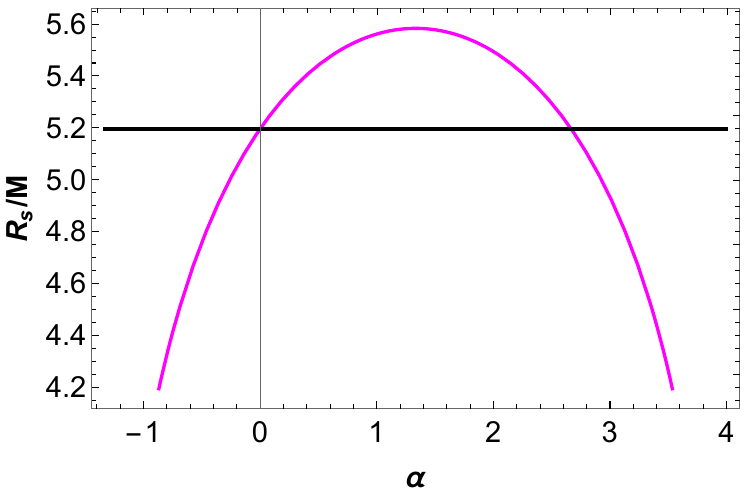}
\caption{Variation of $R_s$ against $\alpha$.
 The horizontal black line corresponds to $R_s=3\sqrt{3}M$. Here $M=1$.}
\label{radius}
\end{figure}

Validating theoretical models against observational or
experimental data requires constraining a free parameter. The
precise bounds on the parameter help ensure the physical viability
and alignment with known phenomena. In this context, estimating
the range of Lorentz violation parameter $\a$ parameter is
important and useful. One can evaluate the consistency of models
such as metric affine bumblebee gravity with data, such as QPOs,
lensing, or QMNs, by, for instance, restricting the Lorentz
violation parameter to a lower limit. By eliminating or endorsing
particular parameter ranges, this procedure improves the
prediction ability of a theory. Here we will be using constraints
reported in \cite{M871, M872, keck, vlti1, vlti2} obtained from
the experimental observation of shadows of supermassive BHs
$M87^*$ and $Sgr A^*$ consistent with
 To this end, we first write the Lagrangian for the
 metric under consideration:
\ba\nn
\mathscr{L}&=&\f{1}{2}\lt(g_{tt} \dot{t}^2+g_{rr}\dot{r}^2+g_{\phi\phi} \dot{\phi}^2\rt)\\
&=&\f{1}{2}\lt(-\fr \dot{t}^2+\f{\dot{r}^2}{\gr}+\hr
\dot{\phi}^2\rt).\label{lagr} \ea Here, we have considered that
the motion is confined in the equatorial plane $\theta=\pi/2$.
 Owing to the static and spherically symmetric nature of spacetime,
 the Lagrangian is independent of time and azimuthal angle. This leads
 to two conserved quantities of motion. These  are as follows:
\ba\nn
\e&=&-p_{t}=-\f{\pt \mathscr{L}}{\pt \dot{t}}=\fr \dot{t},\\\nn
 \l&=&p_{\phi}=\f{\pt \mathscr{L}}{\pt \dot{\phi}}=\hr \dot{\phi}.
\label{pt}
\ea
Here, $\e$ is the energy and $\l$ is the angular momentum.
 Four-velocity of a mass-less particle follow the relation $u_{\mu}u^{\mu}=0$
 where $u^{\mu}=\f{d x^{\mu}}{d\tau}$. This in combination with Eq. [\ref{pt}]
 leads to the following radial equation:
\ba
\nn
&&\dot{r}^2+\lt(-\e^2 \f{\gr}{\fr}+\l^2 \f{\gr}{\hr}\rt)=0\\
&\Rightarrow& \dot{r}^2+V_{eff}(r)=0, \label{rdot} \ea where
$V_{eff}=-\e^2 \f{\gr}{\fr}+\l^2 \f{\gr}{\hr}$ is the effective
potential. Imposing conditions \be V_{eff}(r_p)=0, \quad \f{\pt
V_{eff}}{\pt r}|_{r=r_p}=0, \quad \tx{and} \quad \f{\pt^2
V_{eff}}{\pt r^2}|_{r=r_p}<0, \ee on the effective potential we
obtain radius $r_p$ of the unstable spherical orbits leading to
the relation $\fr' \hr=\hr' \fr$ whose solution yields $r_p$. An
interesting observation one can make is the non-appearance of
$\gr$ in the relation. One may therefore conclude that if a
solution of a proposed model does not affect $\fr$ or $\hr$ (e.g.
\c{bm}), then the photon orbit has a radius equal to $3M$ which is
the case for \s BH. In our case, the radius of the photon orbit
also comes out to be $3M$. The impact parameter corresponding to
the photon orbit is called critical impact parameter $b_p$ as
photons with impact parameter $b < b_p$ get swallowed by the BH
and those with $b > b_p$ get deflected from their path but can
reach asymptotic observer. However, photons with $b=b_p$ circle
around BH several times before either getting swallowed by BH or
reaching the observer. The critical impact parameter provides the
radius of shadow $R_s$ as \be
b_p=R_{s}=\f{\l}{\e}=\sqrt{\f{h{r_p}}{f(r_p)}}=\frac{3}{2}
\sqrt{3} \sqrt[4]{-3 \alpha ^2+8 \alpha +16} M. \ee For $\a=0$, we
restore the value for \s BH $R_s=3\sqrt{3}M$. We display
graphically qualitative dependence of the radius of the shadow on the
parameter $\a$ in Fig. [\ref{radius}]. It reveals that initially,
the radius of the shadow increases with $\a$, reaching a maximum
value $5.58363M$ at $\a=\f{4}{3}$ and then starts decreasing.
Another interesting observation one can make from the figure is
that apart from $\a=0$, we have $R_s=3\sqrt{3}M$ for
$\a=\f{8}{3}$.
To employ observations regarding shadows of $M87^*$ and $Sgr A^*$ for
constraining $\a$, we introduce the parameter $\delta$ defined by \c{del}
\be
\delta=\f{R_s}{3\sqrt{3}M}-1=\frac{1}{2} \sqrt[4]{-3 \alpha ^2+8 \alpha +16}-1.
\ee
It is the deviation of the shadow radius from the \s case.
\begin{figure}[H]
\centering
\includegraphics[width=0.4\columnwidth]{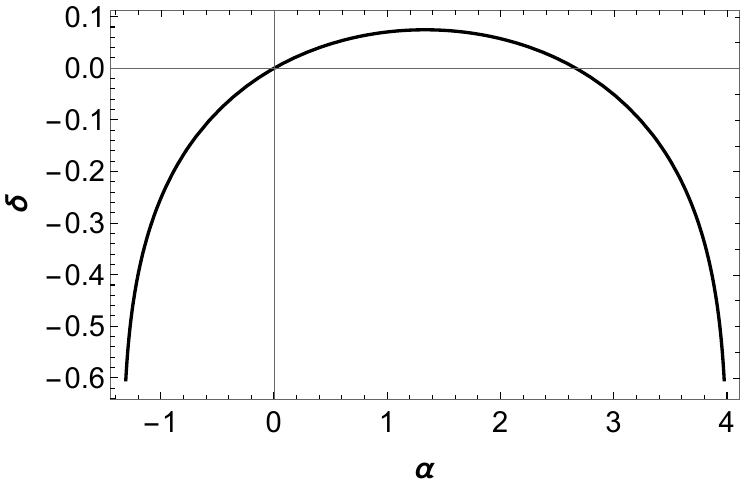}
\caption{Variation of $\delta$ against $\alpha$.}
\label{del}
\end{figure}
Similar to the case of the radius of the shadow  the deviation parameter
too initially increases with $\a$ reaching a maximum value
 $0.0745699$ at $\a=\f{4}{3}$ and then starts decreasing, as evident from
 Fig. [\ref{del}]. Bounds on the deviation parameter are given below
  \cite{M871, M872, keck, vlti1, vlti2}
\begin{center}
\begin{tabular}{|l|c|c|c|r|}
\hline
BH & Observatory & $\delta$ & 1$\sigma$ bounds & 2$\sigma$ bounds\\
\hline
$M87^*$ & EHT & $-0.01^{+0.17}_{-0.17}$ & $4.26\le \frac{R_s}{M}\le 6.03$ &  $3.38\le \frac{R_s}{M}\le 6.91$\\
\hline
\hline
\multirow{2}{*}{$Sgr A^*$}&{VLTI} & $-0.08^{+0.09}_{-0.09}$ &{$4.31\le \frac{R_s}{M}\le 5.25$} &{$3.85\le \frac{R_s}{M}\le 5.72$}\\[3mm]
& Keck & $-0.04^{+0.09}_{-0.10}$ & {$4.47\le \frac{R_s}{M}\le 5.46$} & {$3.95\le \frac{R_s}{M}\le 5.92$}\\[1mm]
\hline
\end{tabular}
\captionof{table}{Bounds on $\delta$ from different observatories.}
\label{bounds}
\end{center}

Subjecting our theoretical prediction to the above bounds, we obtain
the following ranges for the parameter $\a$ that make our model
consistent with the experimental observations.
\begin{center}
\begin{tabular}{|l|c|c|c|c|}
\hline
BH & Observatory & $\delta$ & 1$\sigma$ bounds & 2$\sigma$ bounds\\
\hline
$M87^*$ & EHT & $-0.01^{+0.17}_{-0.17}$ & $[-0.834568, 3.50123]$ &  $[-1.14842, 3.81508]$\\
\hline
\hline
\multirow{2}{*}{$Sgr A^*$}&{VLTI} & $-0.08^{+0.09}_{-0.09}$ &{$[-0.80676, 0.0838442]\cup[2.58282, 3.47343]$} &{$[-1.01439, 3.68106]$}\\[3mm]
& Keck & $-0.04^{+0.09}_{-0.10}$ &{$[-0.714528, 0.540609]\cup[2.12606, 3.38119]$} &{$[-0.975735, 3.6424]$}\\[1mm]
\hline
\end{tabular}
\captionof{table}{Bounds on $\a$ from different observatories.} \label{bounds1}
\end{center}

\begin{figure}[H]
\begin{center}
\begin{tabular}{cc}
\includegraphics[width=0.4\columnwidth]{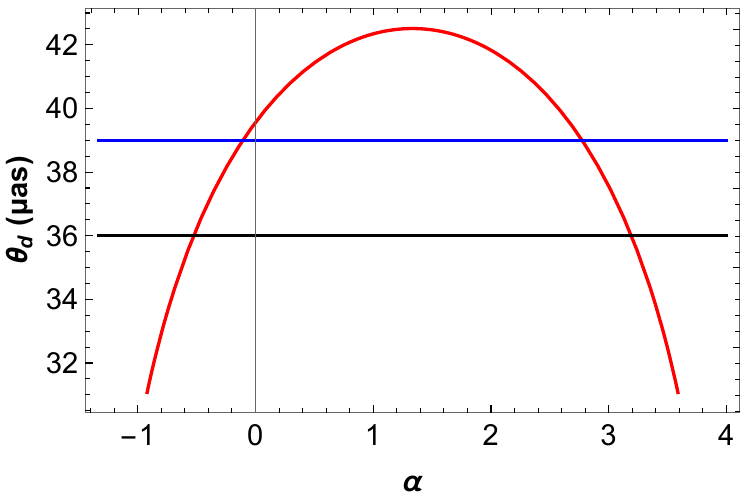}&
\includegraphics[width=0.4\columnwidth]{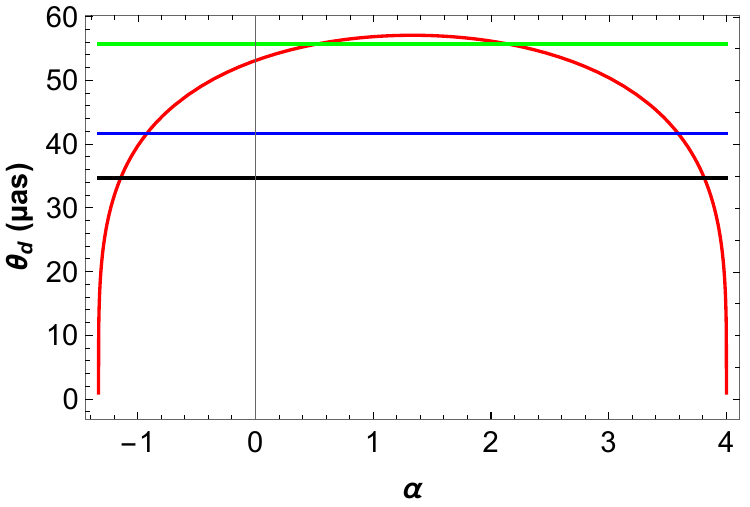}
\end{tabular}
\caption{Variation of angular diameter with $\a$. The left panel
is for $M87^*$ where blue and black horizontal lines correspond to
$\theta_d=39 \mu as$ and $36 \mu as$. The right panel is for $Sgr A^*$
where green, blue, and black horizontal lines are for $55.7 \mu as$, $41.7 \mu as$,
and $34.7 \mu as$, respectively. }\label{angular}
\end{center}
\end{figure}

In addition to bounds on the deviation parameter, we will also use
angular diameter data related to constraining $\a$. The angular
diameter is defined as
\be \theta_d=\f{2R_s}{D},\label{theta}
\ee
where $D$ is the distance of the BH from Earth. According to EHT
collaboration \c{eht, eht4, eht5}, mass, distance, and angular
diameter of  $M87^*$ BH are $M=6.5\pm0.7 \times 10^9 M_\odot$,
$D=16.8\pm0.8 Mpc$, and $\theta_d=42\pm3 \mu as$, respectively.
Those values for $Sgr A^*$ are $M=4.3\pm 0.013 \times 10^6
M_\odot$, $D=8.277\pm 0.033 kpc$, and $\theta_d=48.7\pm7 \mu as$
\c{eht2, vlti1, vlti2}. We use $M=6.5\times 10^9 M_\odot$, $D=16.8
Mpc$ for $M87^*$ and $M=4.3\times 10^6 M_\odot$, $D=8.277 kpc$ for
$Sgr A^*$. We illustrate the variation of angular diameter calculated
using Eq. [\ref{theta}] with $\a$ in Fig. [\ref{angular}].

We can observe from the above figure that even though the angular
diameter for $Sgr A^*$ touches the upper bound in $1\sigma$
confidence level, i.e. $55.7 \mu as$, it never happens for the
$M87^*$ BH. Matching the theoretical prediction and experimental
observations, we obtain $\a \in [-0.106801, 2.77347]$ within
$1\sigma$ confidence level and $\a \in [-0.52517, 3.19184]$ within
$2\sigma$ confidence level for $M87^*$, whereas, for $Sgr A^*$ $\a
\in [-0.921827, 0.5186]\cup [2.14807, 3.58849]$ within $1\sigma$
confidence level and $\a \in [-1.14457, 3.81123]$ within $2\sigma$
confidence level. Our analysis in this section provides a range of
values of the  parameter $\a$  that make the model under
consideration concordant with observed bounds on various
observables related to the shadow of BHs $M87^*$ and $Sgr A^*$. In
the following Sec.  we will be using observed QPOs of microquasars
to constrain the parameter $\a$.
\section{Estimation of the LSB parameter using observational results of QPOs for microquasars} Before delving into details of
QPO, needs to calculate conserved quantities for a test particle in
a time-like geodesics and to obtain effective potential, specific
energy, and angular momentum for a test particle in an equatorial
circular orbit. These quantities are essential in the study of the
epicyclic motion.
\subsection{Motion of the particle in a time-like geodesics}
The Lagrangian for a test particle is given in Eq. [\ref{lagr}]
and its conserved quantities are given in Eq. [\ref{pt}]. However,
in the case of time-like geodesics, $\e$ and $\l$ are specific
energy and specific angular momentum, respectively. The
four-velocity of the test particle follows the relation
$u_{\mu}u^{\mu}=-1$ that provides the radial equation of motion in
the equatorial plane as
\ba\nn \grr
\dot{r}^2&=&-\f{1}{\gtt}\lt[\e^2+\gtt(1+\f{\l^2}{\gth})\rt]\\\nn
&=&-\f{1}{\gtt}\lt[\e^2-\mc{V}_{\text{eff}}\rt]. \ea
 Here,
$\mc{V}_{\text{eff}}=-\gtt(1+\f{\l^2}{\gth})$ is the effective
potential. For motion in a circular orbit of radius $r_0$, we have
\be \veff(r_0)=\e^2, \quad \quad \f{\pt \veff}{\pt r}|_{r=r_0}=0.
\ee The second condition provides the expression for $\l$ and then
we obtain $\e$ using the first condition as
\be
\l^2=\f{-\gtt'\gth^2}{\gth\gtt'-\gtt\gth'},\quad \quad \quad \quad
\e^2 =\f{\gtt^2 \gth'}{\gth\gtt'-\gtt\gth'}, \label{conserved}
\ee
where $'$ represents differentiation with respect to $r$. For
stable circular orbit, we must have $\f{\pt^2 \veff}{\pt
r^2}|_{r=r_0}>0$. The limiting case is the innermost
 circular orbit(ISCO) where we have $\f{\pt^2 \veff}{\pt r^2}|_{r=r_0}=0$ which comes
 out to $6M$ for the metric-affine metric. Thus, similar to the event horizon an
  photon orbit, the ISCO radius too does not depend on $\a$ and matches with that
  for a \s BH. The effective potential $\veff$ and the conserved quantities provide
  the prerequisite platform to study QPOs of microquasars.
\subsection{Epicyclic frequencies}
When a test particle is perturbed from its stable circular orbit of radius $r_0$ in
the equatorial plane, it undergoes epicyclic oscillations, known as quasi-periodic
oscillations. The epicyclic motion has two components: radial component in the
equatorial plane and latitudinal component normal to the equatorial plane. If the
equatorial circular orbit is perturbed by $r=r_0+\delta r$ in the radial direction
 and $\theta=\f{\pi}{2}+\delta \theta$ in the latitudinal direction, where $\delta r$
 and $\delta \theta$ are small quantities, then the differential equations of motion
 governing radial and latitudinal oscillations are
\ba
\delta \ddot{r}+\Omega_{r}^2 \delta r=0,
\quad \quad \delta \ddot{\theta}+\Omega_{\theta}^2 \delta \theta=0,
\ea
where $\Omega_{r}$ and $\Omega_{\theta}$ are locally defined
radial and latitudinal angular frequencies and the dot represents
differentiation with respect to the proper time. To obtain these epicyclic
frequencies, we first separate Hamiltonian into dynamical $\mc{H}_{dyn}$ and $\mc{H}_{pot}$
parts where
\ba\nn
\mc{H}_{dyn}&=&\f{1}{2}\lt(\f{p_{r}^2}{\grr}+\f{p_{\theta}^2}{g_{\theta \theta}}\rt),\\
\mc{H}_{pot}&=&\f{1}{2}\lt(\f{\e^2}{\gtt}+\f{\l^2}{\gth}+1\rt).
\label{hpot} \ea It is the potential part of the Hamiltonian that
governs the local epicyclic oscillations. The epicyclic
frequencies $\Omega_{r}$ and $\Omega_{\theta}$ along with
$\Omega_{\phi}$ are given by \cite{SANJAR, VARBA} \ba\nn
\Omega_{r}^2&=&\f{1}{\grr}\f{\pt^2 \mc{H}_{pot}}{\pt r^2},\\\nn
\Omega_{\theta}^2&=&\f{1}{\grr}\f{\pt^2 \mc{H}_{pot}}{\pt \theta^2},\\
\Omega_{\phi}^2&=&\f{\l}{\gth}.
\label{Omega}
\ea
We must now transform these locally defined angular frequencies to those
measured at spatial infinity. This is done by taking into consideration the
redshift factor. Hence, the required transformation from the locally measured
angular frequencies $\Omega$ to those measured at spatial infinity $\omega$ is
 \c{SANJAR, VARBA}
\be \omega \rightarrow \f{\Omega}{-g^{tt}\e}. \label{omega} \ee
Utilizing expressions for conserved quantities given in Eq.
[\ref{conserved}] along with Eq. [\ref{Omega}] and Eq.
[\ref{omega}], we obtain the following expressions for radial and
latitudinal frequencies in terms of metric coefficients: \ba
\nu_r&=&\f{1}{2\pi} \sqrt{\f{2\gth \gtt'^2-2\gtt \gtt' \gth-\gtt
\gth \gtt''}
{2\gtt \grr \gth}+\f{\gtt' \gth''}{2\grr \gth'}},\\
\nu_\theta&=&\nu_\phi=\f{1}{2\pi}\sqrt{-\f{\gtt'}{\gth'}}.
\label{nu}
\ea
For the metric under consideration, we have
\ba\nn
\gtt&=&-c^2\frac{1-\frac{2 \text{GM}}{c^2 r}}{\sqrt{\left(1-\frac{\alpha }
{4}\right) \left(\frac{3 \alpha }{4}+1\right)}},\\\nn
\grr&=&\frac{\sqrt{\frac{\frac{3 \alpha }{4}+1}{\left(1-\frac{\alpha }{4}\right)^3}}}
{1-\frac{2 \text{GM}}{c^2 r}},\\
\gth&=&r^2.
\ea
Putting the above metric coefficients in Eq. [\ref{nu}] yields
\be
\nu_r=\f{1}{2\pi} \f{c^3}{GM}\sqrt{\frac{(\alpha -4) (2-y)}{(3 \alpha +4) y^4}},
\quad \quad \nu_\theta=\nu_\phi=\f{1}{2\pi} \f{c^3}{GM}
\sqrt{\frac{4}{\sqrt{-3 \alpha ^2+8 \alpha +16} y^3}},
\ee
where $y=r/r_g$, $r_g=\f{GM}{c^2}$, $G$ and $c$ being gravitational
 constant and speed of light, respectively.
\subsection{Constraining the LSB parameter from QPOs of microquasars}
\begin{figure}[H]
\begin{center}
\begin{tabular}{cc}
\includegraphics[width=0.4\columnwidth]{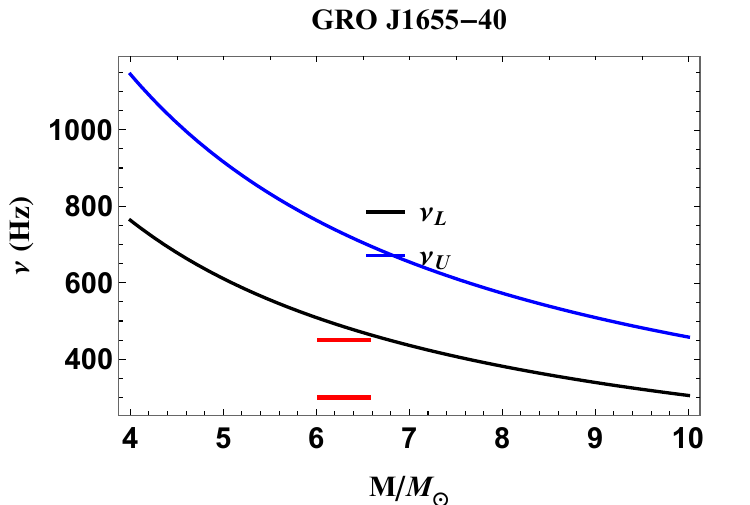}&
\includegraphics[width=0.4\columnwidth]{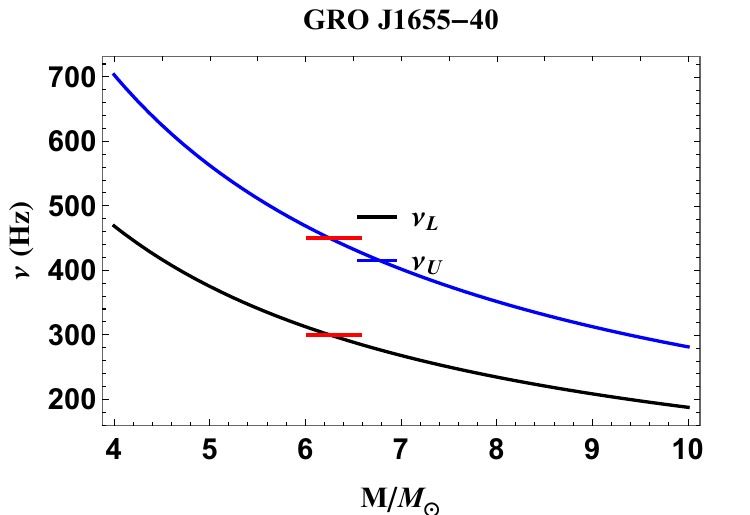}
\end{tabular}
\caption{Fitting the upper and lower frequencies to the observed frequencies
for the GRO J1655-40 microquasar The left one is for $\a=0.2$ and the right
one is for $\a=0.45$. Horizontal lines show the mass error band for GRO J1655-40.}\label{nu40}
\end{center}
\end{figure}

\begin{figure}[H]
\begin{center}
\begin{tabular}{cc}
\includegraphics[width=0.4\columnwidth]{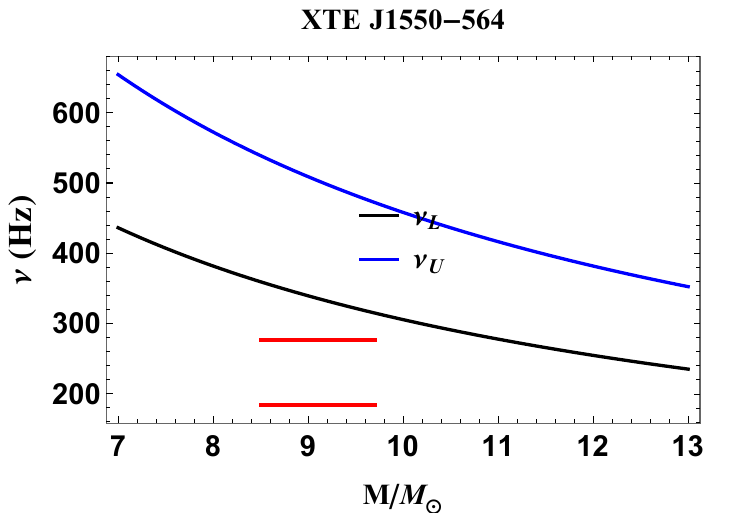}&
\includegraphics[width=0.4\columnwidth]{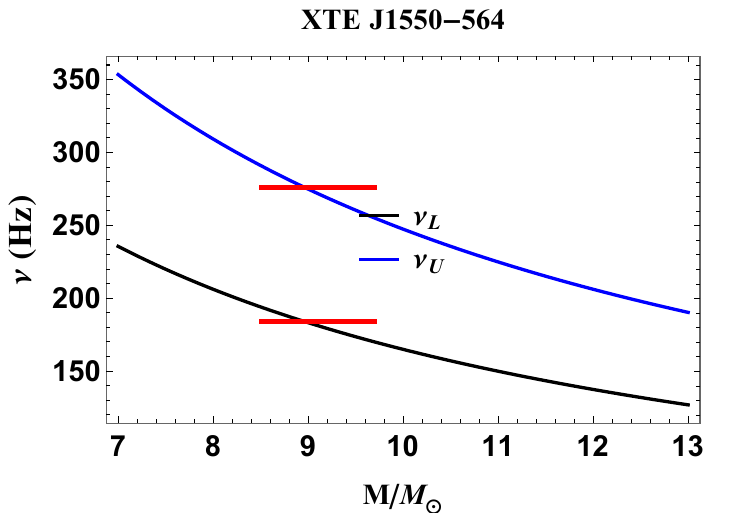}
\end{tabular}
\caption{Fitting the upper and lower frequencies to the observed
frequencies for the XTE J1550-564 microquasar The left one is for $\a=0.2$
and the right one is for $\a=0.5$. Horizontal lines show the mass error band
 for XTE J1550-564.}\label{nu564}
\end{center}
\end{figure}

\begin{figure}[H]
\begin{center}
\begin{tabular}{cccc}
\includegraphics[width=0.4\columnwidth]{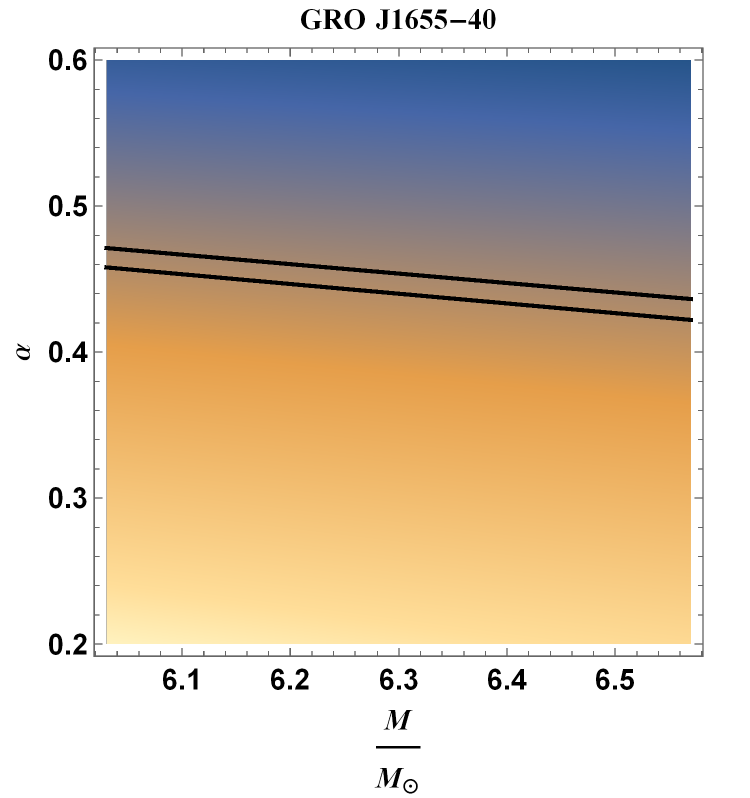}&
\raisebox{.1\height}{\includegraphics[width=0.05\columnwidth]{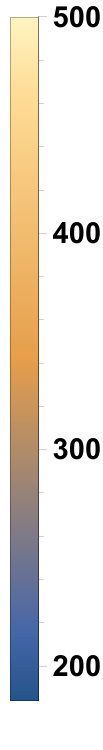}}&
\includegraphics[width=0.4\columnwidth]{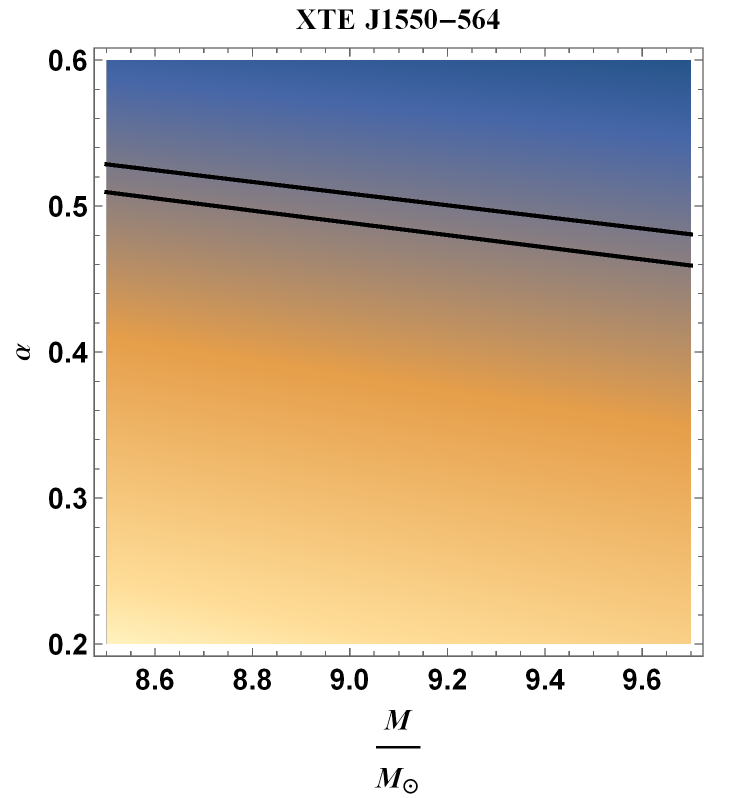}&
\raisebox{.1\height}{\includegraphics[width=0.05\columnwidth]{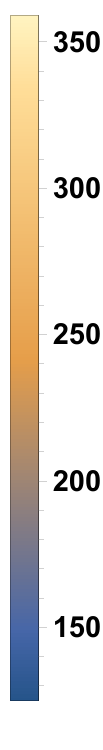}}
\end{tabular}
\caption{Variation of lower QPO frequency $\nu_L$ with mass of the BH and LSB
parameter $\a$. The left one is for GRO J1655-40 and the right one is for XTE J1550-564.
In each plot, the upper solid black line corresponds to the upper $1\sigma$ bound, and
the lower one is for the lower $1 \sigma$ bound of $\nu_L$. }\label{para}
\end{center}
\end{figure}

Apart from the estimation of the LSB parameter from the observation of
shadow of black hole we can exploit observed high-frequency QPOs
(HFQPOs) of microquasars in order to constrain LSB  parameter
$\a$. Now we are in a position to proceed to wads that endeavor.
The process of constraining the parameter $\a$ by utilizing
observed high-frequency QPOs (HFQPOs) of microquasars is indeed
instructive and noteworthy.  The theoretical studies suggest that
the magnitudes of QPOs have a precise dependence on the mass of the BH.
From  recent observations, it is found that the HFQPOs are often
observed in the rational ratio \cite{N32}, especially in the ratio
3:2 \cite{E32},  Two such microquasars which show twin peaks in
their power spectrum in the ratio $3:2$ are $GRO J1655-40$ and
$XTE J1550-564$ galactic microquasars. Their lower ($\nu_{L}$) and
upper ($\nu_U$) QPOs along with their observed masses are as
follows \c{qpo40, qpo401, qpo564}: \ba \tx{GRO J1655-40:}
\f{M}{M_{\odot}}=6.30\pm 0.27,\quad \nu_{U}=450\pm 3 Hz,
 \quad \nu_{L}=300\pm 5 Hz,\\
\tx{XTE J1550-564:} \f{M}{M_{\odot}}=9.10\pm 0.60,\quad
\nu_{U}=276\pm 3 Hz, \quad \nu_{L}=184\pm 5 Hz. \ea Resonance
between radial and vertical oscillations of infilling particles,
especially near ISCO radius, is considered to be one of the viable
explanations for the appearance of twin peaks. This model, known
as resonance model, considers non-linear coupling between the two
oscillations responsible for QPOs \c{resonance, resonance1}. The
frequency ratio $\nu_{U}/\nu_{L}$ for HFQPOs points towards
resonance phenomenon. We will consider the forced resonance model
\c{forced} where \be \nu_{L}=\nu_\theta \quad \quad \tx{and} \quad
\quad \nu_{U}=\nu_{r}+\nu_{\theta}. \ee We are going to employ the
above model in order to bound $\a$. In Fig. [\ref{nu40}]
 and [\ref{nu564}], we fit upper and lower frequencies obtained theoretically
 treating microquasars as LSB BH under consideration to the observed frequencies.
 Frequency curves do not intersect the mass error bands for both the microquasars
  when $\a=0.2$ implying incommensuration of our BH model with observed values
  when $\a=0.2$.
Fig. [\ref{nu40}] and [\ref{nu564}] exhibit the fact that our BH model
is not concordant with observed values of QPOs for all values of $\a$. To
obtain parameter values that make our model commensurate with the experimental
 observations, we provide variation of lower frequency $\nu_L$ as a function
 of mass and LSB parameter in Fig. [\ref{para}].
The region between two solid lines provides the parameter space where observed
values match
theoretical predictions. We obtain the following sets of values of $\a$ from observations:
\ba
&&\tx{For GRO J1655-40: } \quad \a \in [0.422039, 0.471269],\\\nn
&&\tx{For XTE J1550-564: } \quad \a \in [0.459285, 0.528621].
\ea
Interestingly, the above ranges of values do not include $\a=0$ thereby ruling out \s BH as
a viable candidate that may exhibit observed QPO peaks in its power spectrum. Our analysis in
this section makes the claim that our considered model is a feasible candidate that generates
astrophysical
observations commensurate with experimental results stronger.

\section{Concluding remarks}
In this manuscript, we study imprints of LSB, emanating in the
metric-affine bumblebee model,
from the observed shadows of $M87^*$ and $Sgr A^*$ BHs and the observed
QPOs of GRO J1655-40 and
XTE J1550-564 galactic microquasars. These BHs (traceless metric-affine BHs) provide
an excellent
opportunity to probe one of the fundamental pillars of physics, the Lorentz symmetry.
Astrophysical
observations such as shadow and QPOs have little dependence on the complex physics
related to accretion.
 As such, they provide a potent and cleaner tool to probe the nature of the underlying
 spacetime.
 Here we intend to find out the signature of LSB from observables related to shadow and
 QPOs.

We first investigated the impact of LSB on the radius of the unstable photon orbit and the
 corresponding
critical impact parameter. We have found the radius of the photon orbit independent of the
LSB parameter.
Its value came out to be $3M$ which is the value for a \s BH. However, the critical impact
parameter
corresponding to the photon orbit (shadow radius) has a significant dependence on $\a$.
The shadow radius
initially increases with $\a$, reaching a maximum value of $5.58363M$ at $\a=\f{4}{3}$,
and then starts
decreasing. Interestingly, in addition to $\a=0$, the shadow radius for the BH under
 consideration
equals
 to that for a \s BH for $\a=\f{8}{3}$ as well. Introducing the deviation parameter $\delta$,
 we then utilize its observed bounds for $M87^*$ and $Sgr A^*$ BHs reported by EHT, Keck,
  and VLTI to gauge the viability of our BH model. We have tabulated obtained bounds on $\a$
  in Table [\ref{bounds1}]. We have also employed bounds on their angular diameters to
  constrain parameter values of $\a$. Our analysis in this regard has exhibited commensurability
  of our model with observed results for a wide range of values of $\a$.

The motion of test particles in the background of a BH embeds
information regarding background spacetime.
 We have obtained the effective potential for a test particle confined in an equatorial
 circular orbit.
 The ISCO radius is obtained by equating the double derivative of the effective potential
 to zero came out
 to be $6M$, independent of $\a$, and equals that for a \s BH. We have stable circular orbits
 when  $\f{\pt^2 \veff}{\pt r^2}>0$. Thus, stable circular orbits must lie outside ISCO.
 When a particle in a stable circular orbit lying on a latitudinal plane $\Theta$ is
 perturbed
 in radial as well as latitudinal directions, it undergoes epicyclic oscillations in
 two mutually
 perpendicular directions: one in the radial direction in $\Theta$ plane and another in
 latitudinal
  direction normal to the $\Theta$ plane. In this manuscript, we have considered the
   perturbation
  of the equatorial circular orbit.
Observed HFQPOs in the power spectrum of galactic microquasars
  are especially interesting as they mostly occur in the
   rational ratio, especially in the ratio 3:2 \cite{E32}. We have selected two microquasars
  GRO J1655-40 and  XTE J1550-564 with the known QPO data to constrain the LSB parameter.
  The observed
  ratio between lower and higher QPOs points towards resonance between the two epicyclic
  oscillations.
  We have used the forced resonance model. Bound on the parameter $\a$ from QPOs are: for
   GRO J1655-40 $\a \in [0.422039, 0.471269]$ and for XTE J1550-564 $\a \in [0.459285, 0.528621]$.
   These bounds are more stringent than those found in shadow observables. We may obtain
   finer bounds
   on the parameter $\a$ with an improved precision. This we may achieve in the future with
   European
   Space Agency (ESA) X-ray
mission LOFT.


\begin{thebibliography}{the}
\bibitem{COST1} V.A. Kostelecky, R. Potting, CPT, strings, and meson
factories, Phys. Rev. D 51, 3923 (1995).
\bibitem{COST2} D. Colladay, V.A. Kostelecky, CPT violation and the
standard model, Phys. Rev. D 55, 6760 (1997)
\bibitem{COST3}  V.A. Kostelecky, S. Samuel, Spontaneous breaking of Lorentz symmetry in string theory,
 Phys. Rev. D 39, 683 (1989).
\bibitem{COST4}  V.A. Kostelecky, S. Samuel, Gravitational Phenomenology in Higher Dimensional Theories
 and Strings, Phys. Rev. D 40, 1886 (1989).
\bibitem{SAMUEL} V.A. Kostelecky and S. Samuel, Photon and graviton masses in
string theories, Phys. Rev. Lett. 66, 1811 (1991).
\bibitem{eht} Kazunori Akiyama et al. First M87 Event Horizon Telescope Results. I. The Shadow
of the Supermassive Black Hole. Astrophys. J. Lett., 875:L1, 2019.
\bibitem{eht1}Kazunori Akiyama et al. First M87 Event Horizon Telescope Results. IV. Imaging the
Central Supermassive Black Hole. Astrophys. J. Lett., 875(1):L4,
2019.
\bibitem{eht2} Kazunori Akiyama et al. First Sagittarius A* Event Horizon Telescope Results. I. The
Shadow of the Supermassive Black Hole in the Center of the Milky
Way. Astrophys. J. Lett., 930(2):L12, 2022.
\bibitem{eht3} Kazunori Akiyama et al. First Sagittarius A* Event Horizon Telescope Results. II. EHT
and Multiwavelength Observations, Data Processing, and
Calibration. Astrophys. J. Lett., 930(2):L13, 2022.
\bibitem{HAWKING} S. W. Hawking, Commun. Math. Phys. 43, 199 (1975b), [Erratum:
Commun.Math.Phys. 46, 206 (1976)]
\bibitem{BAKE} J. D. Bekenstein, Phys. Rev. D 7 2333  (1973)
\bibitem{REGE} T. Regge, J. A. Wheeler, Stability of a Schwarzschild
singularity  Phys. Rev. 108  1063 (1957)
\bibitem{PRESS} H. W. Press, Long wave trains of gravitational waves
from a vibrating black hole, Astrophys. J 170, L105-L108 (1971)
\bibitem{VISH} V. C.   Vishveshwara, Scattering of gravitational radiation by
 a Schwarzschild black-hole, Nature.  227:936-938 (1970)
\bibitem{KOKKO}  K. D. Kokkotas, B. G. Schmidt, Living Rev. Rel. 2, 2 (1999), arXiv:gr-qc/9909058.
\bibitem{HPN} Hans-Peter Nollert, Class. Quant. Grav. 16, R159 (1999).
Detection of the Schwarzschild precession in the orbit of the star
S2 near the Galactic centre massive black hole. Astron.
Astrophys., 636:L5, 2020.
\bibitem{LIGO} B. P. Abbott et al. GW150914: The Advanced LIGO Detectors in the Era of First
Discoveries. Phys. Rev. Lett., 116(13):131103, 2016.
\bibitem{LENSING1} C. W. Misner, K. S. Thorne, and J. A. Wheeler,
Gravitation [W. H. Freeman and Company, New York, (1971)].
\bibitem{LENSING2} S. Weinberg, Gravitation and Cosmology [John Wiley \& Sons, Inc,
(1972)]
\bibitem{LENSING3} S. Chandrasekhar, The mathematical theory of black holes
[Oxford Classic Texts in the Physical Sciences (1983)]
\bibitem{CASANA} R. Casana, A. Cavalcante, F.P. Poulis,  E.B. Santos,
 Exact Schwarzschild-like solution in a bumblebee gravity model,
\bibitem{KANZI} S. Kanzi and I. Sakalli, GUP Modified Hawking Radiation
in Bumblebee Gravity, Nucl. Phys. B 946, 114703 (2019);
arXiv:1905.00477 [hep-th].
\bibitem{SAKIL} A. Ovgun, K. Jusufi, I. Sakall, Exact traversable
wormhole solution in bumblebee gravity, Phys. Rev. D 99, 024042
(2019); arXiv:1804.09911 [gr-qc].
\bibitem{MALUF} R.V. Maluf, C.A.S. Almeida, R. Casana, and M. Ferreira,
 Einstein-Hilbert graviton modes modified by the Lorentz violating
 bumblebee Field, Phys. Rev. D 90, 025007 (2014); arXiv:1402.3554 [hep-th].
\bibitem{GUIO} J. Paramos and G. Guiomar, Astrophysical
Constraints on the Bumblebee Model, Phys. Rev. D 90, 082002
(2014); arXiv:1409.2022 [astro-ph].
\bibitem {ESCO} C.A. Escobar and A. Martn-Ruiz, Equivalence between bumblebee models
and electrodynamics in a nonlinear gauge, Phys. Rev. D 95, 095006
(2017); arXiv:1703.01171 [hep-th].
\bibitem{COSMOLOGY}  D. Capelo and J. Paramos, Cosmological implications of Bumblebee
vector models, Phys. Rev. D 91, 104007 (2015); arXiv:1501.07685
[gr-qc].
\bibitem{FANG}  W. Liu, X. Fang, J. Jing,  J.
Wang, Exact Kerr-like solution and its shadow in a gravity model
with spontaneous Lorentz symmetry breaking, Eur. Phys. J. C 83, 83
(2023); arXiv:2211.03156 [gr-qc].
\bibitem{MARIZ}  J.F. Assunao, T. Mariz, J.R. Nascimento,  A.Y. Petrov, Dynamical
Lorentz symmetry breaking in a tensor bumblebee model, Phys. Rev.
D 100, 085009 (2019); arXiv:1902.10592 [hep-th].
\bibitem{ADS} A. Uniyal, S. Kanzi, I. Sakall, Greybody factors of bosons
and fermions emitted from higher dimensional dS/AdS black holes in
 Einstein-bumblebee gravity theory, Eur. Phys. J. C 83 668 (2023)  arXiv:2207.10122 [hep-th].
\bibitem{KHODADI} M. Khodadi and M. Schreck, Hubble tension
as a guide for refining the early Universe: Cosmologies with
explicit local Lorentz and diffeomorphism violation, Phys. Dark
Universe 39, 101170 (2023).
\bibitem{GHIL} D. M. Ghilencea, Eur. Phys. J. C 80, 1147 (2020) arXiv:2003.08516
[hep-th]
\bibitem{GHIL1}  D. M. Ghilencea,Palatini quadratic gravity: spontaneous breaking of gauged scale symmetry
and inflation, Eur. Phys. J. C 81, 518 (2021) [arXiv:2007.14733
[hep-th]].
\bibitem{ADEL}  A. Delhom, J. Nascimento, G. J. Olmo, A. Y.  Petrov, P. J. Porfirio,
Metric-affine bumblebee gravity: classical aspects,
Eur.Phys.J. C 81 (2021), 287 arXiv:1911.11605 [hep-th]
\bibitem{ADEL1}  A. Delhom, J. Nascimento, G. J. Olmo, A. Y. Petrov, P.
Porfirio, Radiative corrections in metric-affine bumblebee model,
 Phys. Lett. B 826 (2022) 136932,  arXiv: 2010.06391 [hep-th].
\bibitem{ADEL2}  A. Delhom, T. Mariz, J. R. Nascimento, G. J. Olmo, A. Y. Petrov,
 P. J. Porfirio, Spontaneous Lorentz symmetry breaking and one-loop effective
 action in the metric-affine bumblebee gravity, JCAP 07 (2022)  018 [arXiv:2202.11613 [hep-th]].
\bibitem{AFFINE} A. A. Araujo Filho, J. R. Nascimento, A. Y. Petrov, P. J. PorfiArio:
Vacuum solution within a metric-affine bumblebee gravity
  Phys. Rev. D 108, 085010 (2023),arXiv:2211.11821[gr-qc]
\bibitem{AFFINE1}  G. Lambiase, L. Mastrototaro, Reggie C. Pantig, Ali Ovgun,
Probing Schwarzschild-like Black Holes in Metric-Affine Bumblebee
Gravity with Accretion Disk, Deflection Angle, Greybody Bounds,
and Neutrino Propagation: JCAP 12  026 (2023)
\bibitem{AFFINEOUR}S. K. Jha, A. Rahaman, Nucl.Phys.B 1002 116536
(2024)
\bibitem{QPO1} L. Stella and M. Vietri,
Phys. Rev. Lett. 82, 17-20 (1999) doi:10.1103/PhysRevLett.82.17
[arXiv:astro-ph/9812124 [astro-ph]]
\bibitem{QPO2} L. Stella and M. Vietri, Astrophys. J. Lett. 492, L59 (1998)
doi:10.1086/311075 [arXiv:astro-ph/9709085 [astro-ph]]. 57. C.
GermanA, Phys. Rev. D 98, 083025 (2018). arXiv:1810.12426
[astro-ph.HE]
\bibitem{QPO3}C. Bambi, Phys. Rev. D 85, 043002 (2012).
arXiv:1201.1638 [grqc] 45. C. Bambi, J. Jiang, J.F. Steiner,
Class. QuantumGravity 33, 064001 (2016). arXiv:1511.07587 [gr-qc]
\bibitem{QPO4} A. Tripathi, J. Yan, Y. Yang, Y. Yan, M. Garnham, Y. Yao, S. Li,
Z. Ding, A. B. Abdikamalov, D. Ayzenberg, C. Bambi, T. Dauser,
J.A.Garcia, J. Jiang, S. Nampalliwar arXiv e-prints (2019).
arXiv:1901.03064 [gr-qc]
\bibitem{QPO5} M. Tarnopolski, V. Marchenko, Astrophys. J. 911, 20 (2021).
arXiv:2102.05330 [astro-ph.HE] 59. V.I. Dokuchaev, Y.N. Eroshenko,
Phys. Usp. 58, 772 (2015). arXiv:1512.02943 [astro-ph.HE]
\bibitem{QPO6} M. KoloAs, Z. StuchlAk, A. Tursunov, Class. Quantum Gravity 32, 165009
(2015). arXiv:1506.06799 [gr-qc]
\bibitem{QPO7} A.N. Aliev, G.D.
Esmer, P. Talazan, Class. Quantum Gravity 30, 045010 (2013).
arXiv:1205.2838 [gr-qc]
 \bibitem{QPO8} Z. Stuchlik, P. Slany,G. Torok,
Astron. Astrophys. 470, 401 (2007). arXiv:0704.1252 [astro-ph] 63.
L. Titarchuk, N. Shaposhnikov, Astrophys. J. 626, 298 (2005).
arXiv:astro-ph/0503081
\bibitem{QPO9} J. Rayimbaev, B. Majeed,M. Jamil, K.
Jusufi, A.Wang, Phys.Dark Universe 35, 100930 (2022).
arXiv:2202.11509 [gr-qc]
\bibitem{QPO10} M. Ghasemi-Nodehi, M. Azreg-Anou, K.
Jusufi, M. Jamil, Phys. Rev. D 102, 104032 (2020).
arXiv:2011.02276 [gr-qc]
\bibitem{RKONO} R. A. Konoplya, A. Zhidenko,
 Rev. Mod. Phys. 83, 793 (2011), arXiv:1102.4014 [gr-qc].
\bibitem{CARDOSO}  E. Berti, V. Cardoso,  A. O. Starinets, Class. Quant. Grav. 26, 163001 (2009), arXiv:0905.2975 [gr-qc].
\bibitem{LBAR} L. Barack et al., Class. Quant. Grav. 36, 143001 (2019), arXiv:1806.05195 [gr-qc].
\bibitem{KONO} R. Konoplya, A. Zhidenko, Phys. Lett. B 756, 350 (2016), arXiv:1602.04738 [gr-qc].
[arXiv:2309.13594 [gr-qc]]
\bibitem{M871} Kazunori Akiyama et al.
First Sagittarius $A^*$ Event Horizon Telescope Results. VI.
Testing the Black Hole Metric. Astrophys. J. Lett., 930(2):L17,
2022.
\bibitem{M872} Prashant Kocherlakota et al.
Constraints on black-hole charges with the 2017 EHT observations
of $M87^*$. Phys. Rev. D, 103(10):104047, 2021.
\bibitem{keck} Tuan Do et al.
 Relativistic redshift of the star S0-2 orbiting the Galactic center supermassive black hole. Science, 365(6454):664-668, 2019.
\bibitem{vlti1} R. Abuter et al.
 Mass distribution in the Galactic Center based on interferometric astrometry of multiple stellar orbits. Astron. Astrophys., 657:L12, 2022.
\bibitem{vlti2} R. Abuter et al.
\bibitem{SANJAR} S. Shaymatov, J. Vrba, D. Malafarina, B. Ahmedov,Zdenek Stuchlík, , Phys. Dark Universe 30, 100648 (2020). arXiv:2005.12410 [gr-qc].
\bibitem{VARBA} Z. Stuchlík, , J. Vrba, Eur. Phys. J. Plus 136, 1127 (2021). arXiv:2110.10569 [gr-qc
\bibitem{bm} R. Casana, A. Cavalcante, F. P. Poulis, E. B. Santos, An exact Schwarzschild-like
solution in a bumblebee gravity model    Phys. Rev. D 97, 104001
(2018) [arXiv:1711.02273 [gr-qc]].
\bibitem{del} Kazunori Akiyama et al. First Sagittarius $A^*$ Event Horizon Telescope Results.
\bibitem{eht4}Kazunori Akiyama et al. First M87 Event Horizon Telescope Results. V. Physical
Origin of the Asymmetric Ring. Astrophys. J. Lett., 875(1):L5,
2019.
\bibitem{eht5} Kazunori Akiyama et al. First $M87$ Event Horizon Telescope Results. VI. The Shadow
and Mass of the Central Black Hole. Astrophys. J. Lett.,
875(1):L6, 2019.
\bibitem{resonance} M.A. Abramowicz, V. Karas, W. Kluzniak, W.H. Lee, P. Rebusco,
Publ. Astron. Soc. Jpn. 55, 466 (2003). arXiv:astro-ph/0302183
\bibitem{resonance1} J. Horak, V. Karas, Astron. Astrophys. 451, 377 (2006), arXiv:astro-ph/0601053.
\bibitem{qpo40} M. E. Beer and P. Podsiadlowski, The quiescent light curve and evolutionary state of gro
J1655-40, Mon. Not. Roy. Astron. Soc. 331 (2002) 351,
arXiv:astro-ph/0109136.
\bibitem{qpo401} S. E. Motta et al., Precise mass and spin measurements for a stellar-mass black hole through X-ray timing: the case of GRO J1655-40, Mon. Not. Roy. Astron. Soc. 437 no. 3, (2014) 2554-2565, arXiv:1309.3652 [astro-ph.HE].
\bibitem{qpo564} J. A. Orosz, J. F. Steiner, J. E. McClintock, M. A. P. Torres, R. A. Remillard, C. D. Bailyn, and
J. M. Miller, An Improved Dynamical Model for the Microquasar XTE
J1550-564, Astrophys. J. 730 (2011) 75, arXiv:1101.2499
[astro-ph.SR].
\bibitem{N32}J. E. McClintock and R. A. Remillard. Black hole
binaries, volume 39, pages 157 213. Cambridge, UK: Cambridge
University Press, 2006.
\bibitem{E32}G. T$\ddot{o}$r$\ddot{o}$ k, A. Kotrlova, E. Sramkova,
and Z. Stuchlik. Confronting the models of $3:2$ quasiperiodic
oscillations with the rapid spin of the microquasar GRS $1915+105$
Astron. Astrophys., 531: A59, July 2011. doi: 10.1051/
0004-6361/201015549
\end{thebibliography}
\end{document}